\documentclass[preprint,showpacs,preprintnumbers,amsmath,amssymb]{revtex4}%
\usepackage{graphicx}
\usepackage{dcolumn}
\usepackage{bm}
\usepackage{amsmath}
\usepackage{amsfonts}
\usepackage{amssymb}%
\usepackage{appendix}
\providecommand{\U}[1]{\protect\rule{.1in}{.1in}}
\newcommand{\be}{\begin{equation}}
\newcommand{\en}{\end{equation}}
\newcommand{\bea}{\begin{eqnarray}}
\newcommand{\ena}{\end{eqnarray}}
\begin{document}
\title{G-Warm inflation  }
\author{Ram\'on Herrera}
\email{ramon.herrera@pucv.cl} \affiliation{Instituto de
F\'{\i}sica, Pontificia Universidad Cat\'{o}lica de
Valpara\'{\i}so, Avenida Brasil 2950, Casilla 4059,
Valpara\'{\i}so, Chile.}

\date{\today}

\begin{abstract}
A warm inflationary universe in the context of Galileon model or G-model is studied. Under
a general formalism we study the inflationary dynamics and the cosmological perturbations
considering a coupling of the form $G(\phi,X)=g(\phi)\,X$. As a concrete  example, we
consider an exponential potential together with the cases in which the dissipation and Galilean
coefficients are constants.  Also, we study the weak   regime given by the condition $R<1+3gH\dot{\phi}$, and
the strong regime in which  $1<R+3gH\dot{\phi}$.
Additionally, we obtain constraints on the parameters during the evolution of  G-warm inflation, assuming  the
condition for warm inflation in which the temperature $T>H$, the conditions for the weak and strong
regimes, together with  the consistency relation $r=r(n_s)$ from Planck data.

\end{abstract}

\maketitle

\section{Introduction}

It is well known that in the evolution of the early universe, it presented a
rapid but finite period of the expansion called  inflationary   period or simply
inflation
 \cite{R1,R102,R103,Rin}.
 In this context, inflation
gives an elegant proposal for solving some problems of the
standard  model of the universe;  the flatness, horizon, monopoles,
among other. Nevertheless, an important feature   of the inflationary epoch  is that  inflation
gives account of    the Large-Scale
Structure (LSS) of the universe \cite{R2,R203},  as well as
gives  a causal explanation  of the source of the anisotropies
observed in the Cosmic Microwave Background (CMB)
radiation\cite{astro,Hinshaw:2012aka,Ade:2013zuv,Planck2015,DiValentino:2016foa}.

Among the different models that give account of   the dynamical
evolution  of the inflation, we can mention   the model of  warm
inflation. In the stage of warm inflation the early universe is
filled  with a self-interacting scalar field (inflaton) and the
radiation. In this sense, the model of warm inflation has the
special feature   that it  omits   the reheating stage at the end
of inflation,  since this model  introduces  a decay of a scalar
field  into radiation \cite{warm}. Thus, the scenario of
 warm inflation differs from the  standard inflation or the  cold inflation\cite{Yokoyama:1998ju}. In this form, during
 the stage of
  warm inflation
the decay
takes place  from the dissipative effects, and are caused  from a
friction term introduced on the  background equations. Another important characteristic
 of warm inflation are
the thermal fluctuations produced   from the inflationary epoch. These thermal fluctuations
play  an important    role  in  the initial thermal
fluctuations essential  for the seeds  of the LSS formation \cite{62526,1126}.  As   warm
inflation incorporates in its evolution a radiation field with temperature $T$,
and the  quantum and thermal fluctuations of the scalar field  are directly proportional to the
Hubble parameter $H$ and the temperature, and considering that during
 warm inflation  the thermal fluctuations of the scalar field predominates
over the quantum fluctuations, then the condition
  $T>H$ could be satisfied, see Refs.\cite{warm,62526,Berera:2008ar}.
Also, we mention that  warm inflation ends when the
universe stops inflating and smoothly goes into the radiation era of the
standard
Big-Bang model.
For a  review of models of  warm
inflation model, see Ref.\cite{Berera:2008ar}, and  for
 a representative list of recent references, see Refs.\cite{ac1,ac2}.

On the other hand,  one  general class of inflationary models in which the
expansion of the universe is driven by a minimally coupled scalar field are the
Galilean models of inflation or G-inflation. The model of the G-inflation includes the
canonical and
non-canonical scalar field models of inflation, and in the literature are known
as kinetic gravity braiding models\cite{G1,G2}. However, we can mention that the
most general class of inflationary model corresponds to
 a generalized
G-inflation or G$^2$-inflation that  was studied in Ref.\cite{Kobayashi:2011nu},
here
the authors shown
  the equivalence with the
 Horndeski theory\cite{Ho}.  The
importance of these models is that
  the field equations still include
derivatives only up to second order\cite{Nic}.
The action of the G-model incorporate  an additional term of the form
 $G(\phi,X)\square\phi$ to the standard action of the non canonical scalar field
\cite{G1,G2}. Moreover, the model of G-inflation is an
intriguing type of  inflationary universe, since it has a blue spectrum of the primordial
tensor modes which violates the null energy condition stability\cite{G2,Ka}.
However, considering the slow-roll approximation together with some effective
potentials, these models  cannot generate a blue tensor spectrum\cite{288}.
For example, in  the
model of
Higgs G-inflation that  corresponds to a modification of the standard  Higgs
inflation, in which
  $G(\phi,X)\propto \phi X$,   this situation does not occur \cite{289a}.
  Similarly, in Ref.\cite{289}
 was studied  the case
of the graceful exist  from Higgs G-inflation and also the blue spectrum is prohibited,
see  also Ref.\cite{289} for the generalized Higgs G-inflation.
In particular for the case in which the potential is exactly flat was studied  in G-inflation, this model
was called
 ultra slow roll G-inflation \cite{Ut}, and  here the blue tensor perturbation
 and  the null energy condition are prohibited. The same way,
 an effective potential of the form power law can be found in Ref.\cite{Pw}, and
 depending of the parameters-space the consistency relation could be violated.
Recently, the reheating and the  particle production at the end of inflation in  G-inflation was studied
in  Ref.\cite{Br}, see also\cite{Reh2}, and the difference  from string gas cosmology in Ref.\cite{agr}. Also, in the context of the dark energy
these G-models was studied e.g. in Refs.\cite{DE1,DE2}.

%%%%%%%%%%%%%%%%%%%%%%%%%%%%%%%%%%%%%%%%%%%%%%%

 The goal of this paper is to study the model of the warm inflation in the
 context of
 the
Galilean model or G-model, and  how  a
 dissipative coefficient $\Gamma$  coming from an interaction
 between a standard scalar field and a radiation field influences
the standard  G-cold  model.
   Under a general formalism, we will
 find  the dynamics and  the cosmological perturbations; scalar perturbation and tensor
 perturbation in which the
 function $G(\phi,X)$ is given by $G(\phi,X)=g(\phi)\,X$.  In order to consider
 the model of G-warm inflation at the background level,
 we will analyze the  modified motion of
equations assuming  the slow roll approximation. Also, from the Langevin equation in this framework,  we
will obtain an analytical expression of the power spectrum and its spectral index.
As a concrete example and in order to obtain analytical quantities, we consider an exponential potential, together with a
dissipative  coefficient and Galileon parameter constants. Here, we will consider
 a G-warm inflation  for two regimes, namely the weak and strong regimes, where
 the standard weak and the strong dissipative stages together with the Galileon effect are analyzed.
 In both stages, we will  obtain constraints on the different parameters of our G-warm
 model, considering the condition for warm inflation in which $T>H$,  the
 condition for the weak regime in which $R<1+3gH\dot{\phi}$ (or the strong regime where
 $1<R+3gH\dot{\phi}$), together with the  $r-n_s$ plane from Planck data \cite{Planck2015}.

The outline of the paper is as follows: The next section shows
 the dynamics of the model during the scenario of  G-warm inflation assuming  the
 slow roll approximation.
 In the section III, we find the cosmological perturbations
 specifically,  we obtain explicit expressions for the
 scalar power spectrum, spectral index and tensor to scalar ratio.   In section IV we
analyze a concrete  example for our model,  in which we consider an exponential potential
together with a dissipation
coefficient  constant and a Galileon parameter constant.
Here, we study   the weak and  strong  regimes, respectively.
 Finally, section V resumes our results and exhibits our conclusions. We chose units so that
$c=\hbar=8\pi=1$.

\section{G-warm inflation:background equations}
We begin with
 the 4-dimensional action of the form
\begin{equation}
S = \int \sqrt{-g_{4}}d^{4}x\,\left(\frac{M^{2}_{P}}{2}R
+K(\phi,X)-G(\phi,X)\square\phi
\right)+S_\gamma+S_{int},\label{action}
\end{equation}
where  $g_{4}$  corresponds to the determinant of the space-time
metric $g_{\mu\nu}$, $M_p$ is the Planck mass, $R$ is the Ricci
scalar and $X=-g^{\mu\nu}\partial_{\mu}\phi\partial_{\nu}\phi/2$.
Here, the $K$ and $G$ are arbitrary functions of $X$ and $\phi$, where $\phi$
 corresponds to the scalar
field. Additionally, we consider the radiation action
$S_\gamma$ and the interaction action $S_{int}$. The action
$S_{int}$ represents the interaction of the scalar field with the
other fields\cite{nw1,nw2}.

By considering a spatially  flat  Friedmann Robertson Walker
(FRW) metric, together with  a scalar field homogeneous i.e.,
$\phi=\phi(t)$, the standard   Friedmann equation can be written
as
\begin{equation}
3H^{2} = \kappa\,\rho=\kappa\left[\rho_{\phi_S}
+\rho_\gamma\right],\label{HC}
\end{equation}
where $H=\frac{\dot{a}}{a}$ corresponds to the Hubble parameter,
$a$ is the scalar factor and  $\kappa=1/M_p^2$. Also, here we assume a
two-component system, namely, radiation field, described by an energy density $\rho_\gamma$, and a scalar
field $\phi$ with energy density $\rho_{\phi_S}$.  As said before,
 the universe is filled with a
self-interacting   radiation field and a scalar field.  Also  the total energy density $\rho$ is given by
$\rho=\rho_{\phi_S}+\rho_\gamma$, and the dots denote differentiation with  respect to the time.

 In  this context, from the action given by Eq.(\ref{action}), the energy
density and the pressure associated to the scalar field are\cite{G1,G2}

\begin{equation}
\rho_{\phi_S} = 2K_X\,X-K+3G_XH\dot{\phi}^3-2G_\phi X,\label{r1}
\end{equation}
and
\begin{eqnarray}
p_{\phi_S} = K-2(G_\phi+G_X\ddot{\phi})X,\label{pp1}
\end{eqnarray}
respectively. In the following, we will consider  the subscript $K_X$
corresponds to
 $K_X=\partial K/\partial X$, $G_\phi$ to $G_\phi=\partial G/\partial\phi$,  $K_{XX}=\partial ^2K/\partial X^2$ etc.

%We will assume that the total energy density $\rho$ is confined in the brane,
%and then the continuity equation for the total energy density becomes
%$\dot{\rho}+3\,H\,(\rho+P)=0$.
 From  Ref.\cite{warm} the dynamical equations for
$\rho_{\phi_S}$ and $\rho_{\gamma}$ during  warm inflation
can be written as
\begin{equation}
\dot{\rho_{\phi_S}}+3\,H\,(\rho_{\phi_S}+p_{\phi_S})=-\Gamma\;\;\dot{\phi}^{2},
\label{key_01}%
\end{equation}
and
\begin{equation}
\dot{\rho}_{\gamma}+4H\rho_{\gamma}=\Gamma\dot{\phi}^{2}. \label{key_02}%
\end{equation}
Note that the the continuity equation for the total energy density
becomes $\dot{\rho}+3\,H\,(\rho+p)=0$.

Also, we note that combining Eqs.(\ref{r1}) and (\ref{pp1}), the dynamical
equation for scalar field given by Eq.(\ref{key_01}) can be
rewritten as
$$
  K_X\square\phi+2K_{XX}X\ddot{\phi}+2K_{X\phi}X-K_{\phi}-2(G_\phi-G_{X\phi}X)\square\phi
  +6G_X(\dot{H}X+\dot{X}H
  +3H^2X)
  $$
  \begin{equation}
 +6HG_{XX}X\dot{X} -2G_{\phi\phi}X-4G_{X\phi}X\ddot{\phi}=-\Gamma
  \dot{\phi},\;\;\;\;\mbox{where}\,\,\,\,\;\;\square\phi=\ddot{\phi}+3H\dot{\phi}.\label{eq7}
\end{equation}

In these equations, the coefficient $\Gamma$ denotes the  dissipation  coefficient and
from the second law of thermodynamics the coefficient $\Gamma>0$\cite{warm}. This
coefficient is responsible for the decay of the scalar field into radiation
during the process inflationary of the universe. In general, this
 coefficient  can be defined as
 a function of the temperature of the thermal bath $T$ and
the scalar field $\phi$ i.e., $\Gamma=\Gamma(T,\phi)$. In particular, $\Gamma$
can be expressed
  only  as function the temperature, or
 function the scalar field
 or
merely  a constant \cite{warm,26,28,2802,Zhang:2009ge,BasteroGil:2011xd,BasteroGil:2012cm}.
Also, we mention that for the cases  in which $K=X-V(\phi)$
and $G=0$, we obtained the standard model of warm inflation. Here, $V(\phi)$
corresponds to the effective potential.

In the following we will consider a special case of G-model in which  the functions $K(\phi,X)$ and $G(\phi,X)$
are given by
 \begin{equation}
   K(\phi,X)=X-V(\phi),\;\;\;\;\mbox{and}\;\;\;G(\phi,X)=g(\phi)\,X.\label{anz}
 \end{equation}

 On the other hand, during the  inflationary expansion of the universe,
  the energy density $\rho_{\phi_S}$ predominates  over the density
 $\rho_{\gamma}$ \cite{warm,62526}. At the same time, we consider from Refs.\cite{G1,G2} that
the  effective potential dominates over the quantities $X$, $|G_X
H\dot{\phi}^3|$ and $|G_\phi X|$,
 i.e., $\rho_{\phi_S}\sim V(\phi)$. In this approximation
  the Friedmann equation given by Eq.(\ref{HC}) can be rewritten as
\begin{equation}
3H^{2}\approx \kappa\,\rho_{\phi_S}\approx\,\kappa\,
V(\phi).\label{HH}
\end{equation}

Defining the slow-roll parameters in G-inflation as\cite{G2}
\begin{equation}
\epsilon_1=-\frac{\dot{H}}{H^2},\,\,\,\epsilon_2=-\frac{\ddot{\phi}}{H\dot{\phi}},
\,\,\,\epsilon_3=\frac{g_\phi\dot{\phi}}{gH},\,\,\mbox{and}\,\,\,\epsilon_4=\frac{g_{\phi\phi}X^2}{V_{\phi}},\label{pr}
\end{equation}
and  replacing the functions $K$ and $G$ given by Eq.(\ref{anz})
into Eq.(\ref{eq7}) and considering the slow roll parameters of  Eq.(\ref{pr}) we get
\begin{equation}
3H\dot{\phi}(1+R-\epsilon_2/3-gH\dot{\phi}[3-\epsilon_1-2\epsilon_2-2\epsilon_2\epsilon_3/3])=
-(1-2\epsilon_4)V_\phi,\label{eqf}
\end{equation}
where   $R$ denotes the ratio between the dissipative parameter
$\Gamma$ and the Hubble parameter, and is specified  as
\begin{equation}
R=\frac{\Gamma}{3H}.\label{rG}%
\end{equation}

Assuming  that the slow-roll parameters $\epsilon_1$,
$|\epsilon_2|$, $|\epsilon_3|$, $|\epsilon_4|\ll 1$, then the
slow-roll equation of motion for the field $\phi$ given by
Eq.(\ref{eqf}) results
\begin{equation}
3H\dot{\phi}(1+R+3gH\dot{\phi})\simeq -V_\phi. \label{scalar1}
\end{equation}

Here, we observe different limiting cases.  In the context of the dynamical equations, the limits $R+3gH\dot{\phi}< 1$ and
$1+3gH\dot{\phi}<R$, are the standard weak and strong dissipative regimes in
warm inflation. In the limit $1+R<\mid gH\dot{\phi}\mid $ the Galileon effect predominates, and then
 the
 dynamics
of
warm inflation is modified.  However, there are two interesting limits in which the standard  weak and
strong dissipative regimes are combined with the Galileon effect, and these are the limits  are
$R<1+3gH\dot{\phi}$
and $1< R+3gH\dot{\phi}$, respectively. Here, we call the weak regime to the condition  $R<1+3gH\dot{\phi}$
and strong regime  to $1<R+3gH\dot{\phi}$.

Also, we consider that during warm inflation
 the radiation
production is quasi-stable in which  $\dot{\rho
}_{\gamma}<4H\rho_{\gamma}$ and $\dot{\rho}_{\gamma}<\Gamma\dot{\phi}^{2}%
$, see Refs.\cite{warm,62526}. In this form,   from Eq. (\ref{key_02})   the
energy density of the radiation field can be written as

\begin{equation}
\rho_{\gamma}= C_{\gamma}\,T^{4} \simeq
\frac{\Gamma\,X}{2H},\label{gamm}
\end{equation}
where  $C_{\gamma}%
=\pi^{2}\,g_{\ast}/30$ is a constant,
 and   $g_{\ast}$ denotes  the number of
relativistic degrees of freedom\cite{warm}.

From Eq.(\ref{scalar1}) the rate between the velocity of the scalar field
$\dot{\phi}$ and the Hubble parameter
 results
\begin{equation}
\frac{\dot{\phi}}{H}=-\frac{(1+R)}{2\kappa\,g\,V}\left[1-\left(1-\frac{4\,g\,V_\phi}{(1+R)^2}\right)^{1/2}\right].\label{dphi}
\end{equation}
Here, we choose the negative sign  of the square root, in order to
obtain the limit appropriate of Eq.(\ref{dphi}).
 Thus, taking
 the limit
$g\rightarrow 0$, the  Eq.(\ref{dphi}) coincides with that
corresponding to the standard warm inflation, where
$(\dot{\phi}/H)_{{\lim}_{\;g\rightarrow 0}}=-\frac{V_\phi}{\kappa
V(1+R)}$, see Ref.\cite{warm}.

Also, from Eq.(\ref{pr}) the slow-roll parameter $\epsilon_1$ results
\begin{equation}
\epsilon _1= -\frac{\dot{H}}{H^{2}}\simeq
\frac{(1+R)}{4\kappa\,g}\left(\frac{V_\phi}{V^2}\right)
\left[1-\left(1-\frac{4\,g\,V_\phi}{(1+R)^2}\right)^{1/2}\right].\label{e1}
\end{equation}

Again, we note that in the limit $g\rightarrow 0$, the
Eq.(\ref{e1}) agrees with the standard slow-roll parameter $\epsilon_1$ of warm
inflation.

On the other hand, considering Eqs.(\ref{rG}) and (\ref{e1}), the
 relation between  the energy density of the radiation field
 $\rho_{\gamma}$,
 and
the energy density of the scalar field $\rho_{\phi_S}$, is given
by
\begin{equation}
\rho_{\gamma} \simeq
4\,\kappa\,R\,\epsilon_1^2\,\frac{V^3}{V_\phi^2}\simeq\,4\,\kappa\,R\,
\epsilon_1^2\,\frac{\rho_{\phi_S}^3}{\rho_{\phi_S\,\phi}^2},
\label{eq:rad}
\end{equation}
where we have used that $\rho_{\phi_S}\simeq V$, then $\rho_{\phi_S\,\phi}\simeq
V_\phi$.
Since the inflationary scenario takes place provided that
$\epsilon_1 < 1$ (or equivalently $\ddot{a}>0$) then $
\frac{\rho_{\phi_S}^3}{\rho_{\phi_S\,\phi}^2} >
\frac{\rho_{\gamma}}{4\kappa R}. $

An important quantity corresponds to the number of e-folds $N$ at the end of inflation
that is defined as

\begin{equation}
N = \int^{\phi_{f}}_{\phi_{*}}H  dt\simeq-2\kappa
\int^{\phi_{f}}_{\phi_{*}}
\frac{g\,V}{(1+R)}\,\left[1-\left(1-\frac{4\,g\,V_\phi}{(1+R)^2}\right)^{1/2}\right]^{-1}\,d\phi.\label{Nu}
\end{equation}
In the following,  the subscripts $*$ and $f$ are used  to note
the epoch when the cosmological scales exit the horizon and the
end of inflation, respectively.

\section{Cosmological perturbations}

In this section we will analyze the cosmological perturbations in
our G-warm inflationary model. It is well known that the source of
the density fluctuations correspond to thermal fluctuations during
the scenario of  warm inflation. Therefore, in the evolution of
the warm inflation the fluctuations of the scalar field
$\delta\phi$ are dominantly thermal rather than quantum, see Refs.
\cite{warm,Berera:2008ar}. In this scenario, the curvature and
entropy perturbations are present, since the mixture of the scalar
and radiation fields evolve at the perturbative levels. However,
the entropy perturbations on the large scales decay during warm
inflation and are much smaller that its curvature (adiabatic
modes) counterpart, see
Refs.\cite{warm,Berera:2008ar,DeOliveira:2002wk,cid,Herrera:2006ck}.
In the following, we will study only the curvature perturbations
that can contribute on the large scales. In this context,
following Ref.\cite{warm} the power spectrum of the scalar
perturbation is given by ${\cal{P}_{R}}\simeq \;(H\, /
\dot{\phi})^2\delta \phi^2$, where $\delta\phi$ corresponds to the
thermal fluctuations of the scalar field (see also
Ref.\cite{nonca} the expression of  ${\cal{P}_{R}}$ for the case
of a noncanonical scalar field in warm inflation).

 In order to obtain an analytical
expression for the fluctuations $\delta\phi$,  we need to
calculate the freeze-out wave number $k_F$, since the relation
between $\delta\phi$ and $k_F$ is given by $\delta\phi^2\propto
k_F\,T$\cite{warm}. In this context, we consider
 the Langevin
equation that includes a  thermal stochastic noise $\xi({\bf{k}},t)$. The stochastic noise
satisfies  $<\xi({\bf{k}},t)>=0$, and in the temperature limit
$T\rightarrow \infty$, the stochastic noise is Markovian, such that
$<\xi({\bf{k}},t)\,\xi({\bf{k'}},t)>=2\Gamma
T(2\pi)^3\delta^{3}({\bf{k}}-{\bf{k'}})\delta(t-t')$\cite{ca1,warm,Hall:2003zp}.
Here, the correlation functions can be measured by probability
averages, and  $\bf k$ corresponds
to the wave number vector.

 In this form, assuming  the slow roll scenario and
with the incorporation of the stochastic noise together with the spatial Laplacian term, the
Eq.(\ref{scalar1}) can be rewritten as
\begin{equation}
\frac{d\delta\phi({\bf{k}},t)
}{dt}\approx\frac{1}{(3H+\Gamma+18gH^2\dot{\phi})}\left[-(k^2+V_{\phi\phi})\delta\phi({\bf{k}},t)
+\xi({\bf{k}},t)\right].\label{t1}
\end{equation}
Here, we have considered up to first order in
$\delta\phi({\bf{k}},t)$ and  the Fourier transformation. The approximate
solution of Eq.(\ref{t1}) is given by
\begin{equation}
\delta\phi({\bf{k}},t)\approx
\Theta(k,t)\int_{t_0}^t\,\frac{\xi({\bf{k}},t')}{(3H+\Gamma+18gH^2\dot{\phi})}
\,\Theta^{-1}(k,t')\;dt'+\Theta(k,t)\delta\phi({\bf{k}}e^{-H(t-t_0)},t_0),\label{dels}
\end{equation}
where the quantity $\Theta(k,t)$ is defined as
\begin{equation}
\Theta(k,t)=\exp\left[-\int_{t_0}^t\,\left(\frac{k^2+V_{\phi\phi}}{(3H+
\Gamma+18gH^2\dot{\phi})}\right)\;dt'\right].\label{sold2}
\end{equation}

From Eq.(\ref{dels}) we mention that the first term on the right
hand side realizes the thermalization of  fluctuations
$\delta\phi$ and considers  the efficiency during the thermal  process. The other term
corresponds to  the memory-term that archives the state of the
mode in the beginning $t_0$ of process.  Following
Refs.\cite{warm,Berera:2008ar} during the time interval $\sim
H^{-1}$, and from Eq.(\ref{sold2}), the    freeze-out wave
number is defined by the condition $
\frac{k_F^2}{(3H+\Gamma+18gH^2\dot{\phi})H}=1$. Here, we have
considered that
 the mass term given by $V_{\phi\phi}$ is negligible  in relation to
the  momentum $k^2$.  In this form, the fluctuations of the scalar
field $\delta\phi$ can be written as
\begin{equation}
\delta
\phi^2=\frac{k_F\,T}{2\pi^2}=\frac{\sqrt{3H^2+H\Gamma+18gH^3\dot{\phi}}\;\;T}{2\pi^2}.\label{Fc}
\end{equation}
Note that in the limit $g\rightarrow$ 0, the fluctuations given by  Eq.(\ref{Fc}) reduces
to the fluctuations obtained in Ref.\cite{warm}.

 In this
way, from Eq.(\ref{Fc}) we find that the power spectrum of the
scalar perturbation is given by
\begin{equation}
  {\cal{P}_{R}}\simeq \frac{1}{2\pi^2}\,\left(\frac{H}{\dot{\phi}}\right)^2\,\left[\frac{\Gamma\,X}{2C_\gamma\,H}\right]^{1/4}
   \,\sqrt{3H^2(1+6gH\dot{\phi})+H\Gamma}\;\,\,\,\,.\label{P12}
\end{equation}
Here, we have considered Eq.(\ref{gamm}).

On the other hand, the spectral index $n_s$ is defined as
$n_s=1+\frac{d\ln \,{\cal{P}_{R}}}{d\ln k}$, and  considering Eq.(\ref{P12})
the index $n_s$ becomes
$$
n_s\simeq1-
\frac{\epsilon_1}{2}\left[\frac{7}{2}+\frac{1+12gH\dot{\phi}}{(1+R+6gH\dot{\phi})}\right]
+3\epsilon_2\left[\frac{1}{4}-\frac{gH\dot{\phi}}{(1+R+6gH\dot{\phi})}\right]
+\epsilon_3\left[\frac{3gH\dot{\phi}}{(1+R+6gH\dot{\phi})}\right]
$$
\begin{equation}
+\frac{\epsilon_5}{2}
\left[\frac{1}{2}+\frac{R}{(1+R+6gH\dot{\phi})}\right], \label{ns}
\end{equation}
where the quantity  $\epsilon_5$ is defined as
$$
\epsilon_5=\left(\frac{\dot{\phi}}{H}\right)\,\left(\frac{\Gamma_\phi}{\Gamma}\right),
$$
and the ratio $\frac{\dot{\phi}}{H}$ is given by
Eq.(\ref{dphi}).
%and $\delta_1$ can be seen as a slow-roll parameter

On the other hand, the tensor-perturbation  during inflation would
generate gravitational wave\cite{PT}. In particular during G-warm
inflation the amplitude   of the tensor perturbation is equivalent
to the standard inflationary models
 and the corresponding spectrum
becomes
\begin{equation}
{\cal{P}_{G}}=8\kappa\,\left(\frac{H}{2\pi}\right)^2\approx
\frac{2\kappa^2\,V}{3\pi^2}.\label{onda}
\end{equation}
Here we have used Eq.(\ref{HH}).

A crucial  observational quantity is the tensor-scalar ratio $r$,
which is specified as $r={\cal{P}_{G}}/{\cal{P}_{R}}$. Considering
Eqs.(\ref{P12}) and (\ref{onda}) we find that the tensor to scalar
ratio is given by
\begin{equation}
r=4\kappa\,X\,\left(\frac{2C_\gamma\,H}{\Gamma\,X}\right)^{1/4}\,[3H^2(1+6gH\dot{\phi})+H\Gamma]^{-1/2}\;.\label{te}
\end{equation}

In the following, we will analyze the G-warm inflation for an exponential
potential,  together with a
dissipative coefficient $\Gamma=\Gamma_0$ constant \cite{warm}, and
the Galileon parameter $g=g_0=$ constant\cite{go} .
Also, we will restrict ourselves to the weak and strong dissipative regimes
together with the Galileon effect i.e., $R< 1+3gH\dot{\phi}$ and $1< R+3gH\dot{\phi}$.

\section{The  weak and strong dissipative regimes together with the Galileon effect
 : Exponential potential}

In this section we apply the formalism of above to  G-warm
inflation model, considering  the standard weak and strong dissipative
regimes in which $R=\frac{\Gamma}{3H}< 1$, and $ R>1$, together
with the Galileon effect. In this sense, we call the weak regime
to the condition $R< 1+3gH\dot{\phi}$ and the strong regime to
$1< R+3gH\dot{\phi}$. In order to obtain analytical solutions in
both regimes, we consider the simplest case in which the parameter
$g=g_0=$constant (with units of mass$^{-3}$), and the dissipative
coefficient $\Gamma=\Gamma_0=$ constant (with units of mass).

 For
the specific potential $V(\phi)$, we assume  an exponential
potential defined as
$$
V(\phi)=V_0\,e^{-\alpha\phi},
$$
where $\alpha$ (with units of mass$^{-1}$) and $V_0$ are free
parameters. In the following we shall take $\alpha>0$ and $g_0>0$. We should
mention that this  potential  does not have a minimum,  then the scalar
field does not oscillate around this minimum\cite{RCampuzano:2005qw},  a fundamental requirement for standard
mechanism of reheating in the evolution of cold inflation\cite{Campuzano:2005qw}.
However, during warm inflation there is not separate reheating
scenario as we have mentioned previously, then the problem of
non-oscillating from the exponential potential is not present
here.
%Also we observe that for this potential we have
%$V_\phi=-\alpha V$, then $V_\phi<0$

\subsection{ The weak regime: $R <1+3gH\dot{\phi}$.}

Assuming that the model of G-warm inflation evolves in the regime
in which $R< 1+3gH\dot{\phi}$ i.e., in the weak regime, where the weak dissipative
($R<1$)
and the Galileon effect  dominate the  dynamics  of G-warm inflation.

By considering  Eq.(\ref{dphi}), we find that
$\dot{\phi}$ becomes
\begin{equation}
\dot{\phi}=\frac{1}{2\sqrt{3\,\kappa}\,g_0\,V^{1/2}}\,
\left[\left(1+4\,g_0\alpha\,V
\right)^{1/2}-1\right],\label{wdphi1}
\end{equation}
and assuming the exponential potential, then the solution of
Eq.(\ref{wdphi1}) can be written as
$$
V_0^{-1/2}e^{\alpha \phi/2}(1+\sqrt{1+4\alpha g_0\,V_0
e^{-\alpha\phi}})-2\sqrt{\alpha\,g_0}\;\arcsin(2\sqrt{\alpha\,g_0\,\,V_0
e^{-\alpha\phi}})= \frac{\alpha^2\,}{\sqrt{3\kappa}}\,t+C_1,
$$
where $C_1$ is an integration constant.

The inflationary scenario ends when the universe heats up at
a time when the slow-roll parameter $\epsilon{_1}{_W}\simeq 1$ ( or
equivalently $\ddot{a}\simeq 0$), where the parameter $\epsilon{_1}{_W}$ from
Eq.(\ref{pr})
 is given by
 $$
\epsilon{_1}{_W}\simeq\frac{\alpha}{4\kappa g_0 V}\left[(1+4\alpha\,g_0
 V)^{1/2}-1\right].
 $$
 In this way, the potential at the end of inflation considering
  the condition  $ \epsilon{_1}{_W}(\phi=\phi_f)\simeq 1$,
results
\begin{equation}
V_f=V_0\,e^{-\alpha\phi_f}\simeq
\frac{\alpha}{2\kappa\,g_0}\left[\frac{\alpha^2}{2\kappa}-1\right].\label{wVf}
\end{equation}
Here, we note that as  $V_f>0$, then  we obtain  a lower bound for
the parameter $\alpha$ given by
$\alpha>\sqrt{2\kappa}$.
\begin{figure}[th]
{{\hspace{0cm}\includegraphics[width=3.in,angle=0,clip=true]{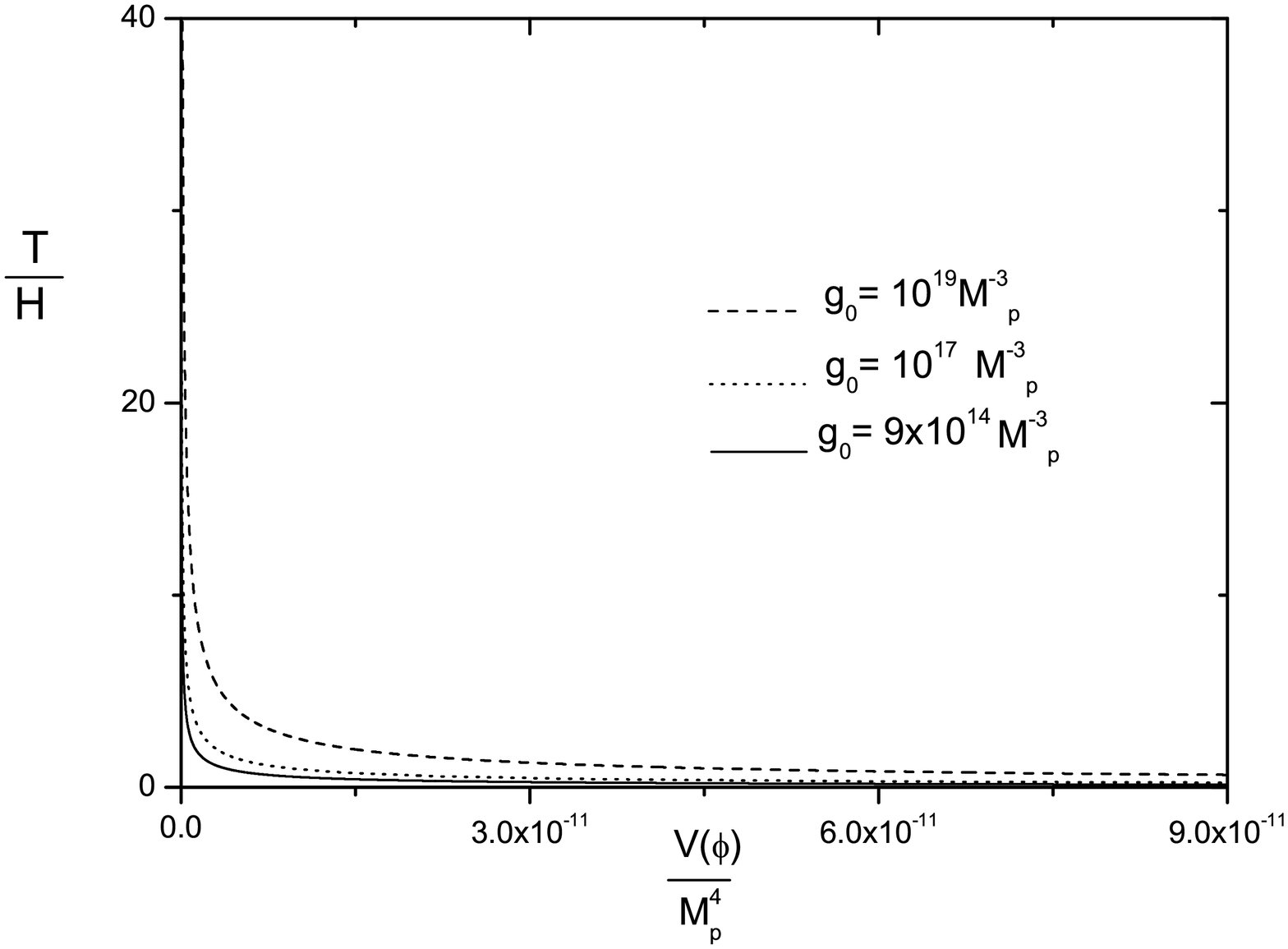}}}
{\includegraphics[width=3.in,angle=0,clip=true]{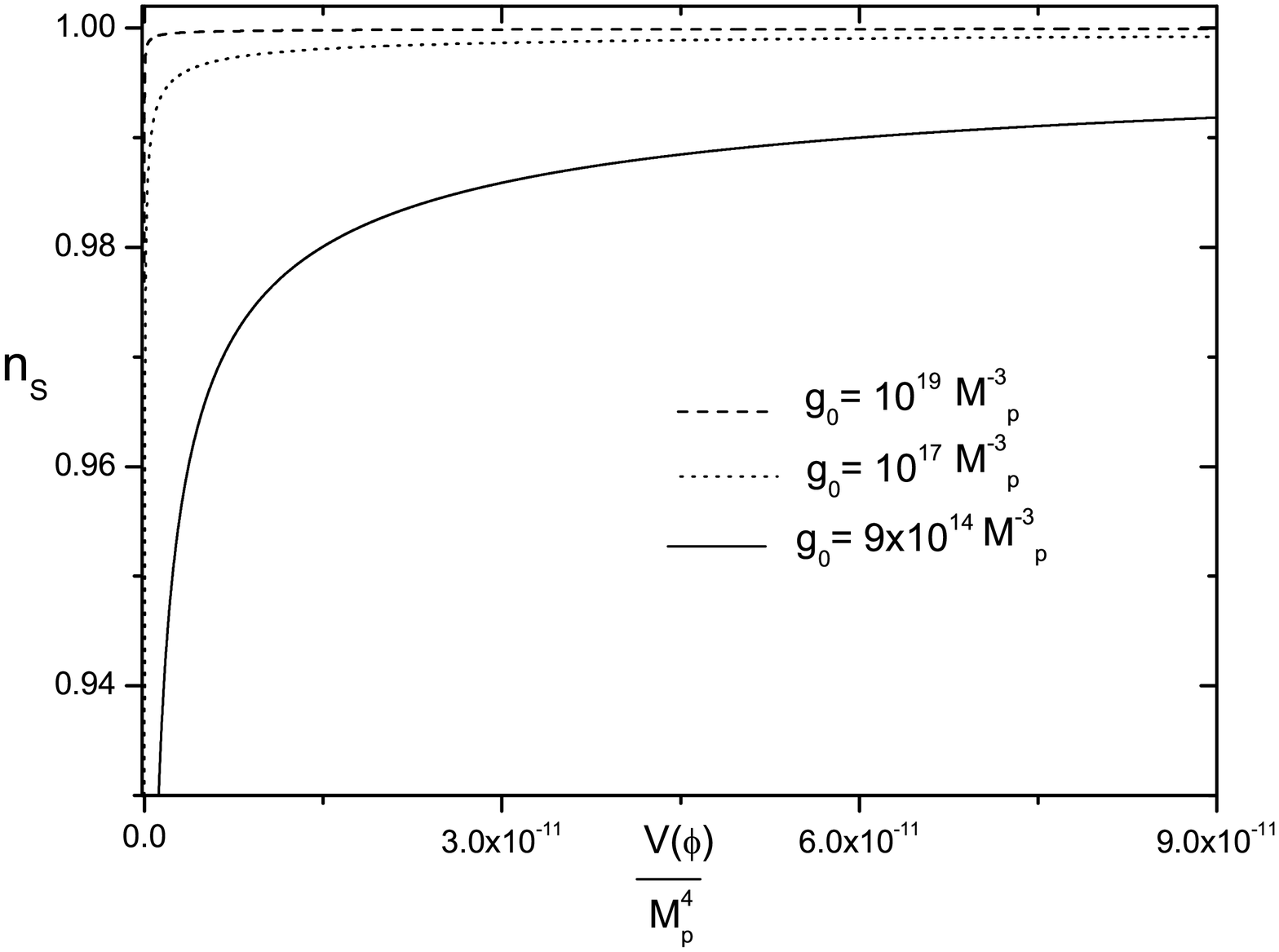}}

{\vspace{-0.5 cm}\caption{ The dependence of the  ratio $T/H$
versus the effective potential $V(\phi)$ (in units of
$M_p^{4}$)(left panel) and the dependence of the the scalar
spectral index $n_s$ versus the potential $V(\phi)$ (right panel)
during the weak regime, for three different values of the
parameter $g_0$ (in units of $M_p^{-3}$).
 In both panels, the dashed, dotted  and
solid lines correspond  to the pairs ($\alpha=1.42M_p^{-1}$,
$\Gamma_0=5.1\times 10^{-6}M_p$), ($\alpha=1.43M_p^{-1}$,
$\Gamma_0=1.1\times10^{-10}M_p$), and ($\alpha=1.42M_p^{-1}$,
$\Gamma_0=1.2\times10^{-12}M_p$), respectively. In these plots we
have used the value $C_\gamma=70$ .
 \label{fig01}}}
\end{figure}

\begin{figure}[th]
{{\hspace{0cm}\includegraphics[width=3.in,angle=0,clip=true]{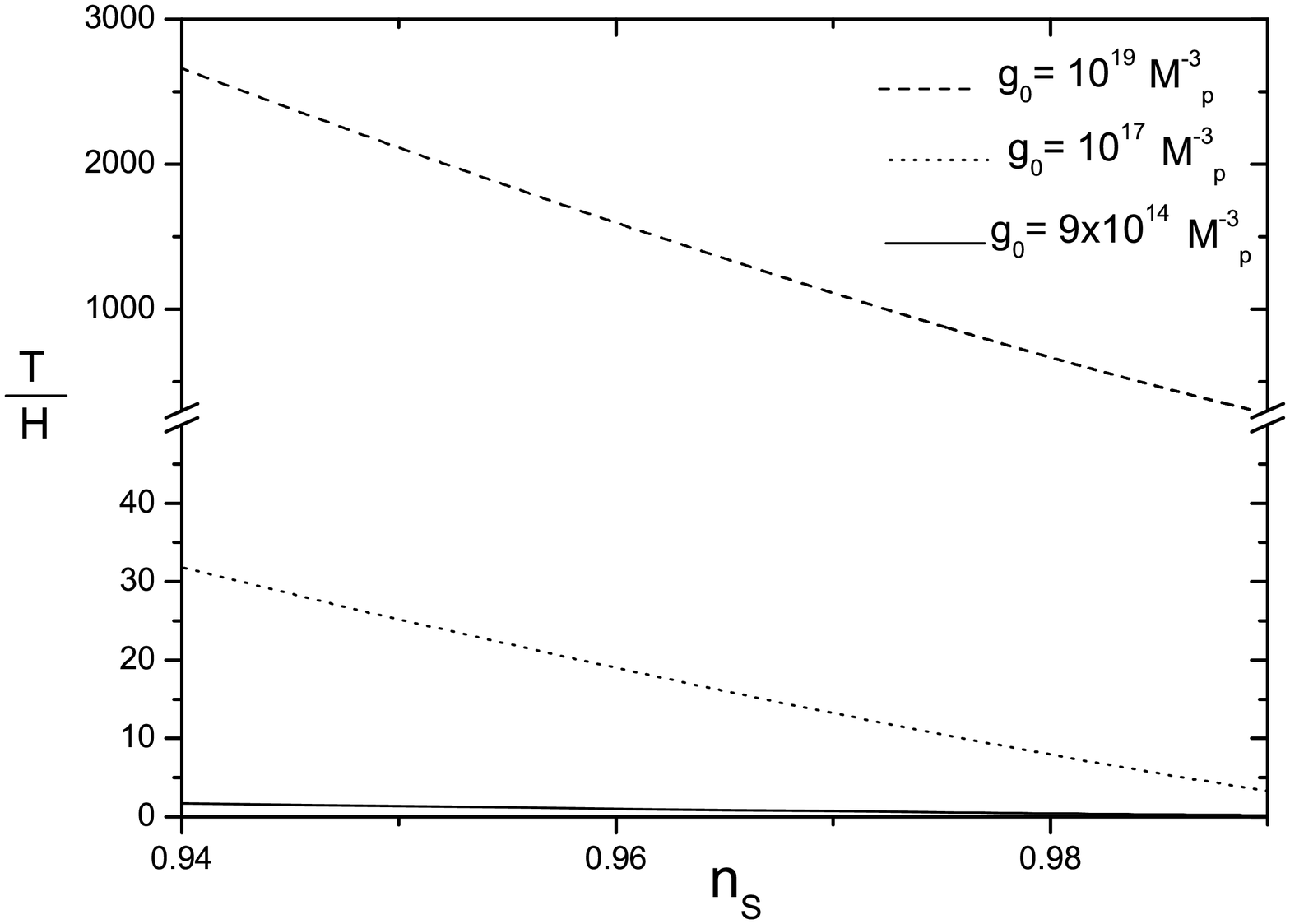}}}
{\includegraphics[width=3.in,angle=0,clip=true]{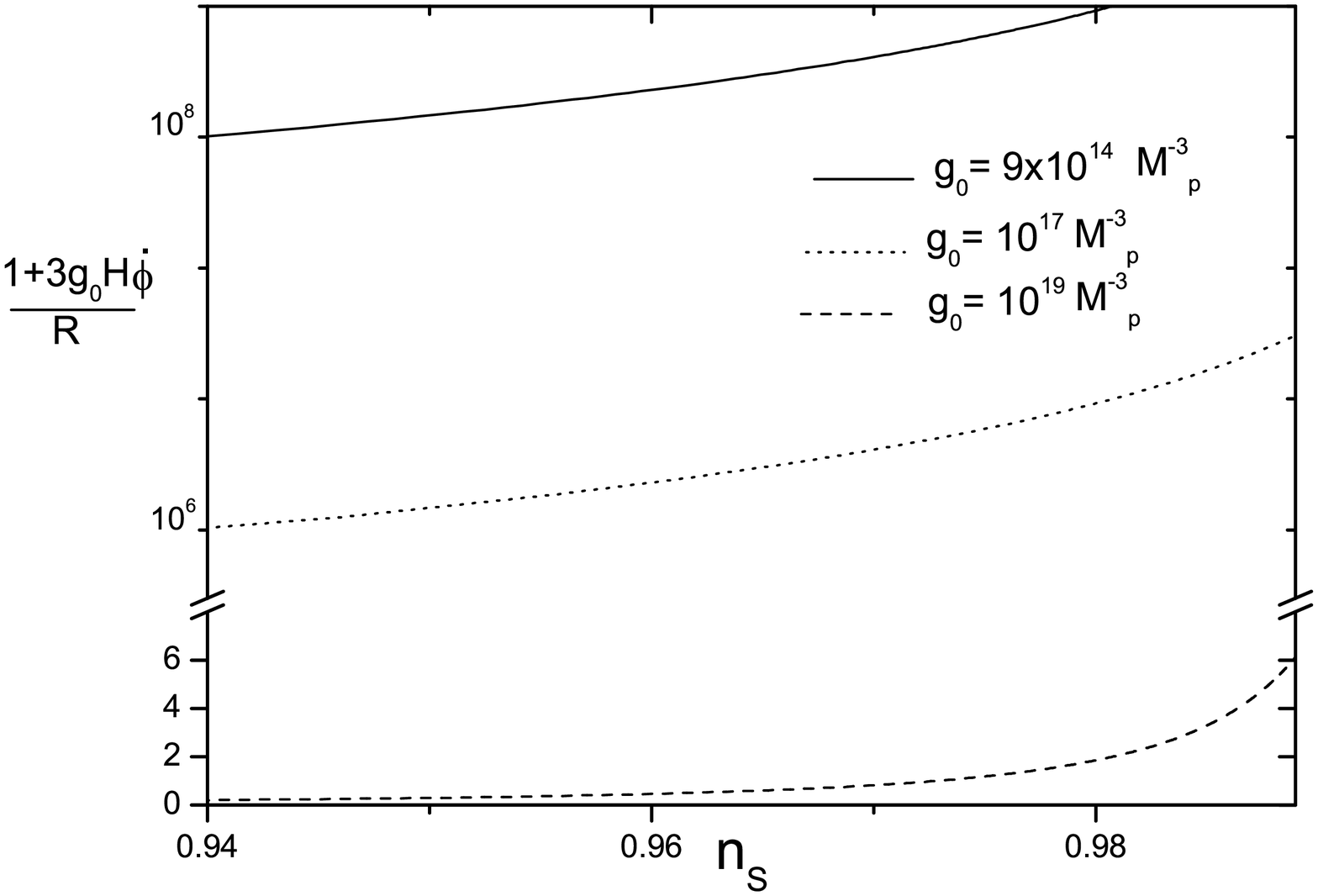}}

{\vspace{-0.5 cm}\caption{ The dependence of the  ratio $T/H$
versus the scalar spectral index $n_s$ (left panel) and the
dependence of the ratio $(1+3g_0H\dot{\phi})/R$ versus the scalar
spectral index $n_s$ (right panel) during  the weak regime, for
three different   values of the parameter $g_0$ (in units of
$M_p^{-3}$).
 In both panels, the dashed, dotted  and
solid lines correspond  to the pairs ($\alpha=1.42M_p^{-1}$,
$\Gamma_0=5.1\times 10^{-6}M_p$), ($\alpha=1.43M_p^{-1}$,
$\Gamma_0=1.1\times10^{-10}M_p$), and
($\alpha=1.42M_p^{-1}$, $\Gamma_0=1.2\times10^{-12}M_p$),
respectively. In these plots we have used the value $C_\gamma=70$
.
 \label{fig1}}}
\end{figure}

\begin{figure}[th]
%{{\hspace{0cm}\includegraphics[width=3.in,angle=0,clip=true]{figure1.eps}}}
{\hspace{-4.0cm}\includegraphics[width=3.in,angle=0,clip=true]{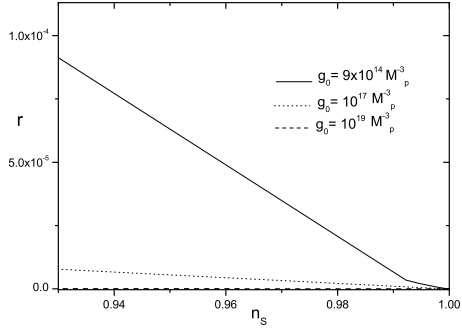}}
{\vspace{-0.5 cm}\caption{ The  consistency relation $r=r(n_s)$
during the weak regime, for three different values of the
parameter $g_0$ (in units of $M_p^{-3}$).
 As before,
 the dashed, dotted  and
solid lines correspond  to the pairs ($\alpha=1.42M_p^{-1}$,
$\Gamma_0=5.1\times 10^{-6}M_p$), ($\alpha=1.43M_p^{-1}$,
$\Gamma_0=1.1\times10^{-10}M_p$), and ($\alpha=1.42M_p^{-1}$,
$\Gamma_0=1.2\times10^{-12}M_p$), respectively. In this plot we
have used the value $C_\gamma=70$.
 \label{fig2a}}}
\end{figure}

During this regime, the number of e-folds at the end of inflation becomes
\begin{equation}
N=\frac{\kappa}{\alpha^2}\,\left[\sqrt{1+4\alpha g_0V_*}-
\sqrt{1+4\alpha g_0V_f} + \ln\left(\frac{\sqrt{1+4\alpha
g_0V_*}-1}{\sqrt{1+4\alpha g_0V_f}-1}\right)\right],\label{Nw}
\end{equation}
where $V_f$ is given by Eq.(\ref{wVf}). Here we have used that $d\phi=dV/V_\phi=-\alpha^{-1}d(\ln
V)$ for the exponential potential.

On the other hand, from Eq.(\ref{P12}) the power spectrum of the scalar perturbation
${\cal{P}_{R}}$ during this scenario results
\begin{equation}
{\cal{P}_{R}}\simeq\frac{2g_0^2}{\pi^2}\frac{\kappa^{5/2}\,V^{5/2}}{[(1+4g_0\alpha
V)^{1/2}-1]^2}\,\left[\frac{\sqrt{3}\,\Gamma_0 X}{2C_\gamma
\sqrt{\kappa V}}\right]^{1/4}\,(1+2g_0\sqrt{3\kappa
V}\,\dot{\phi})^{1/2},\label{Pw}
\end{equation}
where $\dot{\phi}$ in terms of the potential $V$ (or scalar field
$\phi$, since $V=V_0 e^{-\alpha \phi}$) is given by
Eq.(\ref{wdphi1}).

The spectral index $n_s$ from Eq.(\ref{ns}) is given by
\begin{equation}
n_s\simeq1-\frac{3\epsilon{_1}{_W}}{4}\left[\frac{3+22g_0\sqrt{3\kappa V}\dot{\phi}/3}
{1+2g_0\sqrt{3\kappa V}\dot{\phi}}\right]+\frac{3\epsilon{_2}{_W}}{4}\left[\frac{1+2g_0\sqrt{3\kappa V}\dot{\phi}/3}
{1+2g_0\sqrt{3\kappa V}\dot{\phi}}\right],\label{nsw}
\end{equation}
where the slow parameter $\epsilon{{_2}{_W}}$ during the weak
scenario is defined as
$$
\epsilon{_2}{_W}\simeq\,\left[\frac{\alpha}{\kappa V(1+2
\sqrt{3\kappa V}\dot{\phi})}\right]\; \left[\alpha V-\frac{\sqrt{3\kappa V}\dot{\phi} }{2}
-6 \kappa g_0 V \dot{\phi}^2\right].
$$

From Eq.(\ref{te}) we obtain that the tensor to scalar ratio during the weak regime can be written as
\begin{equation}
r\simeq
\left(\frac{2C_\gamma\,\sqrt{\kappa V}}{\sqrt{3}\,\Gamma_0}\right)^{1/4}\,
\left(\frac{16\,\kappa\,X^{3/2}\,}{ V\,(1+2g_0\sqrt{3\kappa V}\,\dot{\phi})}\right)^{1/2}.
\end{equation}

 In Fig.\ref{fig01} we show the dependence  of the rate $T/H$
and the spectral index $n_s$ on the exponential potential
$V(\phi)=V_0e^{-\alpha\phi}$ (in units of $M_p^{4}$) during the
 regime in which $R< 1+3g_0H\dot{\phi}$ (weak regime). In both plots we have
used three different values of the parameter $g_0$ (in units of
$M_p^{-3}$). The left panel shows the essential condition for warm
inflation in which the temperature of the thermal bath $T>H$ in
terms of the exponential potential $V(\phi)$ (or  the scalar
field). The right panel shows the dependence of the spectral index
$n_s$ in terms of the exponential potential $V(\phi)$.  In order
to write down the ratio $T/H$ and the index $n_s$ as a function of
the potential $V(\phi)$  during this regime, we consider
Eqs.(\ref{HH}), (\ref{gamm}), and (\ref{wdphi1}), and then we
obtain the rate  between the temperature of the thermal bath $T$
and the Hubble parameter $H$, as a function of the exponential
potential $V(\phi)$. Analogously, we consider
Eqs.(\ref{wdphi1})and (\ref{nsw}), and together with the slow roll
parameters $\epsilon{_1}{_W}$ and $\epsilon{_2}{_W}$, we find the
curve of the spectral index $n_s$ as a function of the effective
potential $V(\phi)$. In both panels we have used  the value of
$C_\gamma=70$. From the left panel, we note that the essential
condition for warm inflation $T>H$ occurs for the values of the
potential $V< 9\times 10^{-11}M_p^4$ for the curve  in which
$g_0=10^{19}M_p^{-3}$. Analogously, the condition $T>H$ occurs for
$V<4\times10^{-11}M_p^4$ in the case $g_0=10^{17}M_p^{-3}$, and
$V<4\times10^{-12}M_p^4$ for the curve in which
$g_0=9\times10^{14}M_p^{-3}$. In the right panel, we observe that
the spectral index $n_s=n_s(\phi)$  is always $n_s\lesssim1$.
Also, we note that the index $n_s$ takes the observational value
$n_s=0.967$ for the value of the potential $V\simeq 5\times
10^{-16}M_p^4$ for the curve in which $g_0=10^{19}M_p^{-3}$,
$V\simeq 5\times 10^{-14}M_p^4$ in the case in which
$g_0=10^{17}M_p^{-3}$, and $V\simeq 6\times 10^{-12}M_p^4$ in the
case in which $g_0=9\times10^{14}M_p^{-3}$.

On the other hand, in order to obtain  the parameter-pair
($\alpha, \Gamma_0$) for a given value of the parameter $g_0$, we
consider the observational data in which $n_s\simeq 0.967$ and
${\cal{P}_{R}}\simeq 2.2\times 10^{-9}$ from Planck experiment
data, and also we assume that the number of e-folds $N=60$. In
this form, we numerically consider Eqs.(\ref{Nw}), (\ref{Pw}) and
(\ref{nsw}) and we find that the parameter $\alpha\simeq
1.42M_p^{-1}$ and $\Gamma_0\simeq 5.1\times 10^{-6}M_p$ for the
value of the parameter $g_0=10^{19}M_p^{-3}$ for which $n_s\simeq
0.967$, ${\cal{P}_{R}}\simeq 2.2\times 10^{-9}$ and $N=60$.
Analogously,   for the value $g_0=10^{17}M_p^{-3}$ corresponds to
$\alpha\simeq1.43M_p^{-1} $ and $\Gamma_0\simeq 1.1\times
10^{-10}M_p $ and for the case in which
 $g_0=9\times 10^{14}M_p^{-3}$ we
have $\alpha\simeq 1.42 M_p^{-1}$ and $\Gamma_0\simeq1.2\times
10^{-12}M_p$.

 In the following we will analyze the plots of the essential
condition for warm inflation, and the condition of weak (or
strong) regime in terms of the spectral index $n_s$. In this form,
we study the conditions in our model with the help of  the
observational data of the spectral index $n_s$, since it has a
very well defined range from the marginalized constraints Planck
data (68$\%$ and 95$\%$ confidence levels)\cite{Planck2015}.

 In Fig.\ref{fig1} we show the
dependence of the rates $T/H$ and $(1+3g_0H\dot{\phi})/R$ on the
spectral index $n_s$ during the weak regime i.e., $R<
1+3g_0H\dot{\phi}$. In both plots we have used three different
values of the parameter $g_0$ (in units of $M_p^{-3}$). The left
panel shows the essential condition for warm inflation in which
the temperature of the thermal bath $T>H$. Here, we confirm that
the condition $T>H$ takes place. The right panel shows  the
dependence of the dimensionless quantity $(1+3g_0 H\dot{\phi})/R$
during inflation and   we check that the model evolves in the weak
regime in warm inflation. In order to write down the rates $T/H$
and $(1+3g_0H\dot{\phi})/R$ in terms of the scalar spectral index
$n_s$ during this regime, we take into account Eqs.(\ref{HH}),
(\ref{gamm}), (\ref{wdphi1}) and (\ref{nsw}),  and we numerically
obtain the parametric plot for the rate  between the  temperature
of the thermal bath $T$ and the Hubble parameter $H$, as a
function of the scalar spectral index $n_s$. Similarly, we use
Eqs. (\ref{HH}), (\ref{wdphi1}) and (\ref{nsw}), and we
numerically find the parametric plot of the curve
$(1+3g_0H\dot{\phi})/R$  as a function of the spectral index
$n_s$.  In both panels we have considered the value of
$C_\gamma=70$.

In Fig.\ref{fig2a} we show the  consistency relation $r=r(n_s)$
during the weak regime, for three different values of the
parameter $g_0$.  In particular, for the value of tensor to scalar
ratio $r=r(n_s=0.967)$, we  find that for the value
$g_0=10^{19}M_p^{-3}$ (together with the pair $\alpha\simeq
1.42M_p^{-1}$ and $\Gamma_0\simeq 5.1\times 10^{-6}M_p$ ) the
ratio $r=r(n_s=0.967)\simeq2.32\times10^{-8}$, for the value
$g_0=10^{17}M_p^{-3}$ corresponds to
 $r=r(n_s=0.967)\simeq1.95\times 10^{-6}$ and for the case in which  $g_0=9\times 10^{14}M_p^{-3}$
 corresponds to
 $r=r(n_s=0.967)\simeq3.85\times 10^{-5}$. Thus, for the different values of $g_0$ (and the pairs
 ($\alpha$,$\Gamma_0$)),  we find that the tensor to scalar ratio are
 well corroborated from Planck data where $r<0.1$. Also,  we observe that
 the values of the consistency relation $r=r(n_s)\sim
 0$.
 In this form, we note that the tensor to
 scalar ratio gives  $r\sim 0$, and we conclude that the consistency relation $r=r(n_s)$ does not
 add a new constraint on the parameter $g_0$ during this stage.

 In this form, from left panel of the  Fig.\ref{fig1}, we find
  a lower bound for $g_0>9\times 10^{14}M_p^{-3}$, considering the essential condition for the scenario
  of warm inflation takes place i.e., $T/H>1$. Now from the right panel, we obtain
  an upper bound for the parameter $g_0<10^{19}
 M_p^{-3}$  from the condition that the model evolves in the weak regime in
which $R<(1+3g_0H\dot{\phi})$. In this way, we achieve that the interval of the
parameter $g_0$ during the weak regime is given by
$9\times10^{14}<g_0\,M_p^3<10^{19} $. The same form
 for the dissipative parameter
$\Gamma_0$ we find that the range
is given by $1.2\times 10^{-2}<\Gamma_0 M_p^{-1}<5.1\times 10^{-6}$ and for the parameter $\alpha$
a value very close to
$\alpha\sim \sqrt{2\kappa}$ (see condition of Eq.(\ref{wVf})).

\subsection{ The strong regime: $1< R+3gH\dot{\phi}$.}

Considering  that the model of G-warm inflation develops  in the
strong regime
 in which the strong dissipative regime together with the Galileon effect ($1< R+3gH\dot{\phi}$)
 predominate the
 evolution of warm inflation,
 we find that the
$\dot{\phi}$ can be written as
\begin{equation}
\dot{\phi}=\frac{\Gamma_0}{6\kappa\,g_0\,V}\,\left[\left(1+
\frac{12\kappa\,g_0\alpha\,V^2}{\Gamma_0^2}\right)^{1/2}-1\right],\label{dphi1}
\end{equation}
and the solution of the scalar field in terms of the cosmological
time can be written as
$$
e^{\alpha\phi}[1+(1+b_1\,e^{-2\alpha\phi})^{1/2}]+\sqrt{b_1}{\mbox
{
arcsinh}}(\sqrt{b_1}\,e^{-\alpha\phi})=\frac{2\alpha^2\,V_0}{\Gamma_0}\,\,\left[\,t+C_1\,\right],
$$
where the dimensionless constant $b_1$ is defined as
$b_1=\frac{12\alpha \,g_0\kappa V_0^2}{\Gamma_0^2}$, and $C_1$ is an
 integration constant. Here we have considered that $\dot{\phi}=\frac{\dot{V}}{V_\phi}=-\frac{\dot{V}}{\alpha
 V}$.

As before, assuming  that the inflationary scenario ends when the universe heats up at
a time when the slow-roll parameter $\epsilon_{1_{S}}\simeq 1$, where now the
parameter $\epsilon_{1_{S}}$ from Eq.(\ref{pr}) is
defined as
$$
\epsilon_{1_{S}}\simeq\frac{\alpha \Gamma_0}{4\sqrt{3}g_0
\kappa^{3/2}
V^{3/2}}\left[\left(1+\frac{12g_0\alpha\kappa\,V^2}{\Gamma_0^2}\right)-1\right],
 $$
 during this regime. Thus, under the  condition $\epsilon_{1_{S}}\simeq 1$, we find that
 the real solution for
  the potential at the end
of inflation $V_f$ can be written as
\begin{equation}
V_f=V_0\,e^{-\alpha\phi_f}\simeq\frac{2A^2\,B}{3}+\frac{A^4\,B^2}{3D}+\frac{D}{3},\label{Vf}
\end{equation}
where the constants $A$, $B$ and $D$ are defined as
$$
A=\frac{\alpha\Gamma_0}{4\sqrt{3}\,\kappa^{3/2}\,g_0},\,\,\,B=\frac{b_1}{V_0^2}=\frac{12\alpha
\,g_0\kappa}{\Gamma_0^2},\,\,\,\mbox{and}\,\,\,\,D=[54A^2+A^6B^3+6A^2\sqrt{3(27+A^4B^3)}]^{1/3},
$$
respectively.

Also, during this scenario the number of e-folds $N$ is given by
\begin{equation}
N=\frac{1}{A\,B^{3/4}}\,
[\Xi(V_f)-\Xi(V_*)],\label{NfS}
\end{equation}
where the function $\Xi(V)$ is defined as
$$
\Xi(V)=\frac{1+\sqrt{1+BV^2}-B^{1/4}V^{1/2}\;\sqrt{1+BV^2}\,\,
_2F_{1}[1/4,1/2,3/2,1+BV^2]}{B^{1/4}V^{1/2}},
$$
and $_{2}F_1$ is the hypergeometric function\cite{Libro}. Here,
$V_f$ is given by Eq.(\ref{Vf}) and also we have used that
$d\phi=dV/V_\phi=-\alpha^{-1}d(\ln V)$.

On the other hand,  considering that the fluctuations of the scalar field  can be
written as $\delta\phi^2=
(1/2\pi^2)\left[\frac{\sqrt{3}\Gamma_0\,X}{2C_\gamma\sqrt{\kappa
V}}\right]^{1/4}\,\sqrt{H_*(\Gamma_0+6\kappa g_0 V
\dot{\phi})}$ during this stage, then   the power spectrum of the
scalar perturbation ${\cal{P}_{R}}$, from Eq.(\ref{P12})  becomes
\begin{equation}
{\cal{P}_{R}}\simeq\,\frac{2\,\times
3^{7/8}}{\pi^2}\left(\frac{g_0}{\Gamma_0[(1+BV^2)^{1/2}-1]}\right)^2
\left[\frac{\Gamma_0 X(\kappa
V)^{25/2}}{2C_\gamma}\right]^{1/4}[\Gamma_0+6\kappa g_0V\dot{\phi}]^{1/2
},\label{PrS}
\end{equation}
where
$\dot{\phi}=\dot{\phi}(V)$ is given by Eq.(\ref{dphi1}).

From Eq.(\ref{ns}) the scalar spectral index $n_s$ during this
scenario results
\begin{equation}
n_{s}\simeq 1-\frac{\epsilon_{1_{S}}}{4}\left[\frac{7R+22g_0\sqrt{3\kappa V}\,\dot{\phi}}
{R+2g_0\sqrt{3\kappa V}\,\dot{\phi}}\right]+\frac{3\epsilon_{2_{S}}}{4}
\left[\frac{R-2g_0\sqrt{3\kappa V}\dot{\phi}/3}{R+2g_0\sqrt{3\kappa
V}\dot{\phi}}\right],
\label{ns1S}
\end{equation}
where the  slow-roll parameter $\epsilon_2$, in the strong regime
$\epsilon_{2_{S}}$ (for $
\Gamma+9gH^2\dot{\phi}>3H$) becomes
$$
\epsilon_{2_{S}}\simeq\frac{\sqrt{3\,V_*}\;\alpha}
{\sqrt{\kappa}}\,\,\left[\frac{\alpha-3\kappa
g_0\dot{\phi}_*^2}{\Gamma_0+6\kappa g_0V_* \dot{\phi}_*}\right].
$$

\begin{figure}[th]
{{\hspace{0cm}\includegraphics[width=3.in,angle=0,clip=true]{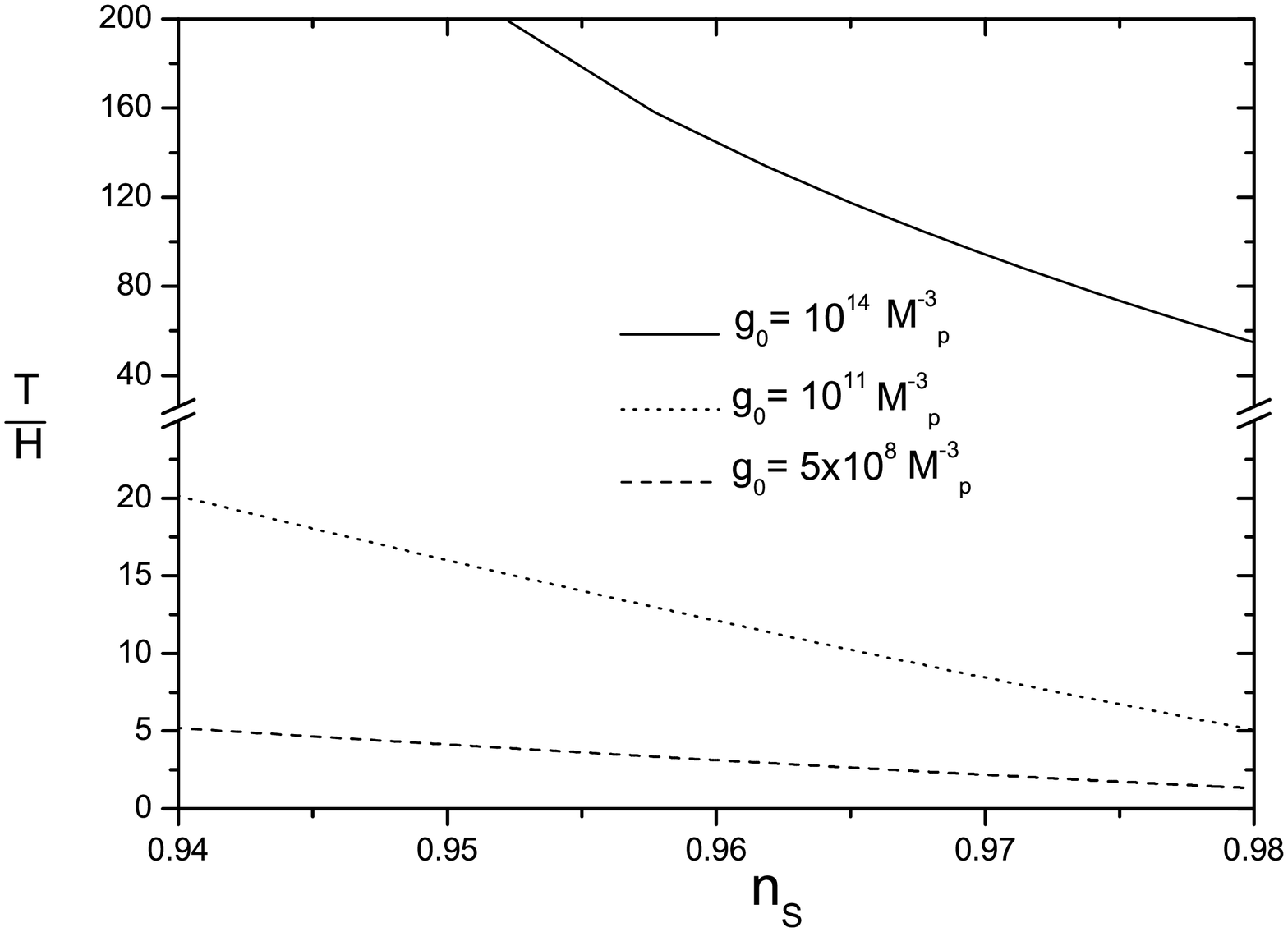}}}
{\includegraphics[width=3.in,angle=0,clip=true]{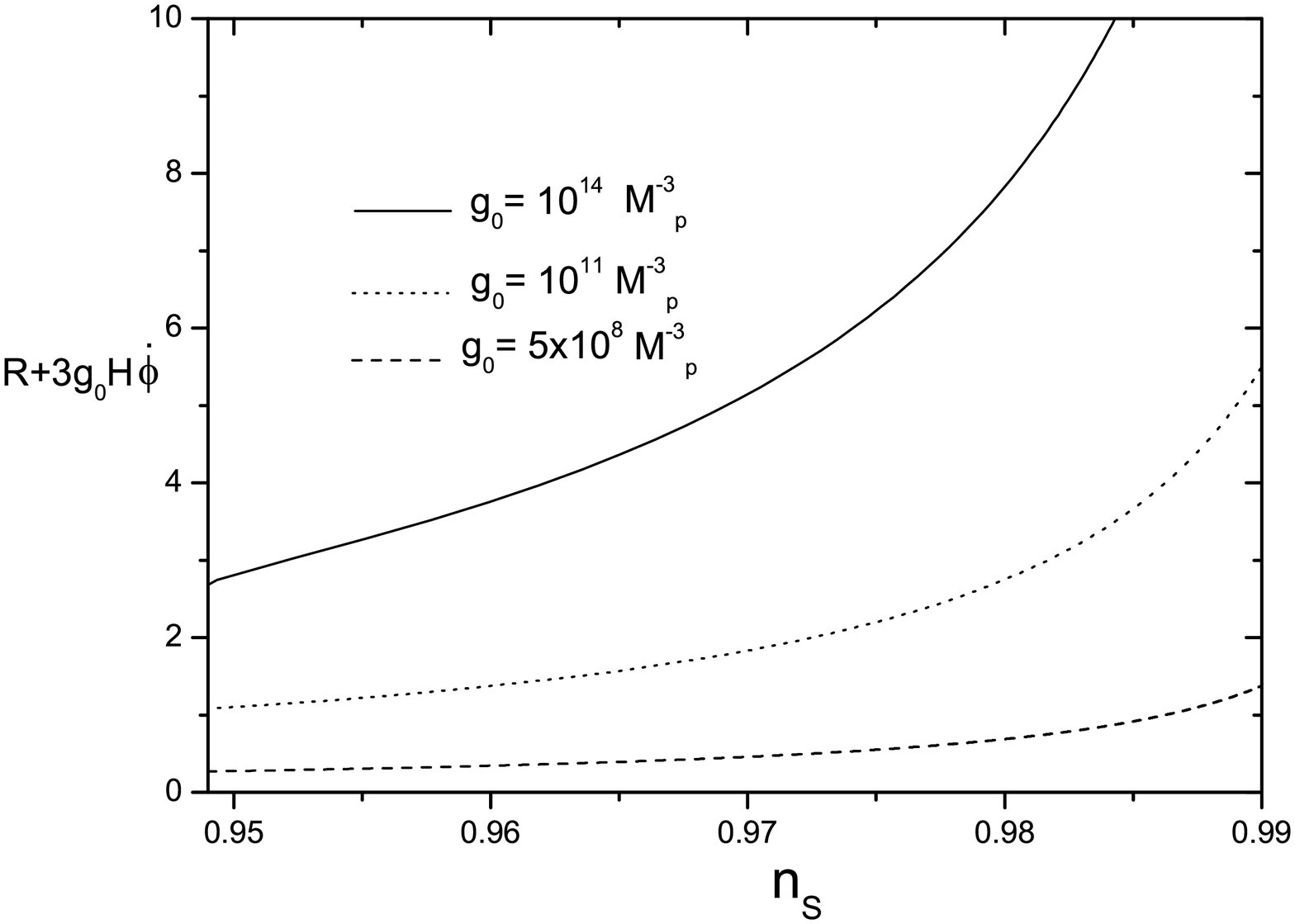}}

{\vspace{-0.5 cm}\caption{ Left panel:  the ratio $T/H$ versus the
scalar spectral index $n_s$. Right panel: the quantity
$(R+3g_0H\dot{\phi})$ versus the scalar spectral index $n_s$. As before,
for both panels we have considered three different values of the
parameter $g_0$ (in units of $M_p^{-3}$), during  the strong  regime in which $1<
R+3gH\dot{\phi}$. In both plots, the solid, dotted and dashed
lines correspond to the pairs ($\alpha\simeq0.34M_p^{-1}$, $\Gamma_0\simeq6.1\times
10^{-7}M_p$), ($\alpha\simeq0.21M_p^{-1}$, $\Gamma_0\simeq
5.1\times 10^{-8}M_p$), and($\alpha\simeq0.11M_p^{-1}$,
$\Gamma_0\simeq6.9\times10^{-9}M_p$),
respectively. In these plots we have used the value $C_\gamma=70$.
 \label{fig2}}}
\end{figure}

For the tensor-to-scalar ratio $r={\cal{P}_{G}}/{\cal{P}_{R}}$,
 we have
\begin{equation}
r\simeq \left[\frac{2^9\times\sqrt{3}\; C_\gamma\,\kappa^{7/2}\; X^3}{\Gamma_0\,\sqrt{V}
}\right]^{1/4}\;\;\left[\Gamma_0+6\kappa\,g_0\,V\,\dot{\phi}\right]^{-1/2}.\label{rS}
\end{equation}
%\begin{equation}
%\times\left[\frac{\sqrt{3}\Gamma_0\,X_*}{2C_\gamma\sqrt{\kappa
%V_*}}\right]^{1/4}\,\sqrt{\Gamma_0-6\kappa g_0 V_*
%\dot{\phi}_*}\;\;\left(\sqrt{1-b_1V_*^2}\right)^{-\frac{12\alpha\kappa\,g_0}{b_1\Gamma_0^2}}\;V_*^{\frac{3}{2}-\frac{b_1\Gamma_0^2}{16\alpha\kappa\,g_0}}\;.
%\end{equation

Here we have used Eq.(\ref{te}).

In Fig.\ref{fig2} we show the evolution of the rates $T/H$ (left
panel) and $R+3g_0H\dot{\phi}$ (right panel) on the spectral index
$n_s$ during the strong regime in which $1< R+3g_0H\dot{\phi}$. As
before, we take into account three values of the parameter $g_0$
(in units of $M_p^{-3}$). Considering Eqs.(\ref{gamm}),
(\ref{dphi1}) and (\ref{ns1S}) we numerically obtain the
parametric plot for the rate $\frac{T}{H}=\frac{T}{H}(n_s)$ and
the quantity $[R+3g_0H\dot{\phi}]=[R+3g_0H\dot{\phi}](n_s)$ during
warm inflation. In addition, as before from Eqs.(\ref{NfS})
(\ref{PrS}) and (\ref{ns1S}) we numerically  find that the
parameter $\alpha\simeq0.34M_p^{-1}$ and $\Gamma_0\simeq6.1\times
10^{-7}M_p$ for the value of the parameter $g_0=10^{14}M_p^{-3}$
for which the spectral index $n_s=0.967$,
${\cal{P}_{R}}=2.2\times10^{-9}$ and the number of e-folds $N=60$.
Similarly, for the value of the parameter $g_0=10^{11}M_p^{-3}$
gives the values $\alpha\simeq0.21M_p^{-1}$ and $\Gamma_0\simeq
5.1\times 10^{-8}M_p$ and for the case in which $g_0=5\times
10^{8}M_p^{-3}$ corresponds to $\alpha\simeq0.11M_p^{-1}$ and
$\Gamma_0\simeq6.9\times10^{-9}M_p$, respectively. Thus, in the plots of
 Fig.\ref{fig2} the solid, dotted and dashed lines correspond to
the pairs ($\alpha\simeq0.34M_p^{-1}$, $\Gamma_0\simeq6.1\times
10^{-7}M_p$), ($\alpha\simeq0.21M_p^{-1}$, $\Gamma_0\simeq
5.1\times 10^{-8}M_p$), and($\alpha\simeq0.11M_p^{-1}$,
$\Gamma_0\simeq6.9\times10^{-9}M_p$)), for the values of
$g_0=10^{14}M_p^{-3}$, $g_0=10^{11}M_p^{-3}$ and $g_0=5\times
10^{8}M_p^{-3}$, respectively. As before in these plots we have
used the value $C_\gamma=70$.

On the other hand, we consider Eqs.(\ref{ns1S}) and (\ref{rS}) and we
numerically obtain the parametric relationship  of the consistency relation $r=r(n_s)$ during the strong regime.
For the value of the parameter $g_0=10^{14}M_p^{-3}$
 (together with the pair $(\alpha\simeq0.34M_p^{-1},\Gamma_0\simeq6.1\times 10^{-7}M_p)$) we numerically find
 that the tensor to scalar ratio is $r=r(n_s=0.967)\simeq3.73\times 10^{-5}$.
 For the case in which the parameter $g_0=10^{11}M_p^{-3}$ we obtain
  $r=r(n_s=0.967)\simeq2.34\times 10^{-3}$ and for the value $5\times 10^{8}M_p^{-3}$ corresponds
  to $r=r(n_s=0.967)\simeq0.42$ and this value is disproved from observational data, since $r<0.1$ (Planck satellite).
   Thus,  we note that for the values $g_0=10^{14}M_p^{-3}$ and
   $g_0=10^{11}M_p^{-3}$,
   the tensor to scalar ratio during the strong regime becomes
  $r\sim 0$,  and these values of $r$ are well corroborated with the Planck data (figure not shown).

 In  this way, from the left panel of Fig.3 we observe that for value of the
  parameter $g_0>5\times 10^{8}M_p^{-3}$ (a lower bound) is well supported by the
  essential condition of warm inflation $T>H$, but is disapproved by Planck data. However, from the right
  panel we also obtain  a lower bound for the parameter $g_0>10^{11}M_p^{-3}$ from the
  condition $1< R+3g_0H\dot{\phi}$ i.e., the strong regime. Thus, we only can
  get  a lower limit for the parameter $g_0$ given by $10^{11}<g_0
  M_p^{3}$. Similarly, we find that for the parameters $\alpha$ and $\Gamma_0$
  the limits are given by $\alpha>0.21M_p^{-1}$ and $\Gamma_0>5.1\times 10^{-8}
  M_p$, respectively.
   Additionally, we numerically observe  that for values of the parameter
   $g_0>10^{11}M_p^{-3}$,
  the tensor to scalar ratio $r\rightarrow 0$.

\section{Conclusions \label{conclu}}

In this paper we have investigated  the model of G-inflation in the framework  of the warm inflation.
Under a general formalism  we have found the dynamics and the
cosmological perturbations in our G-warm inflationary model in the
context of  the slow roll approximation. In this general analysis
we have obtained from the Langevin equation the fluctuations of
the scalar field $\delta\phi$ in order to obtain  an analytical
expression for the power spectrum. As a concrete example we have
considered an exponential potential and we have applied our
results considering the standard weak and strong dissipative
regimes together with the Galileon effect. Also,  in order to
obtain analytical solutions in both regimes, we have consider the
simplest case in which the parameter $g=g_0=$constant and the
dissipative coefficient $\Gamma=\Gamma_0=$ constant. From these
parameters we have obtained analytical quantities  in the slow
roll approximation for the corresponding scalar field, number of
e-folds, power spectrum, spectral index and tensor to scalar
ratio. Thus,  from these expressions and in both scenarios we have found
the constraints on the parameters in our model. In this sense, in
both regimes, we have obtained constraints for the parameters considering
the condition of warm inflation in which the temperature of the
thermal bath $T>H$; from the conditions of the weak and strong
regime where $R<1+3gH\dot{\phi}$ and $1<R+3gH\dot{\phi}$, and also from
the Planck data in which we have considered the constraint on the
consistency relation $r=r(n_s)$.

In our analysis for the both regimes,  we have numerically found
the pair ($\alpha, \Gamma_0$) for a given value of the parameter
$g_0$, from the  observational Planck-data in which the spectral index
$n_s\simeq 0.967$, ${\cal{P}_{R}}\simeq 2.2\times 10^{-9}$, and also we have considered
the number of e-folds $N=60$.

For the weak regime in which $R<1+3gH\dot{\phi}$ we have found the
constraints on the parameters $g_0, \alpha$ and $\Gamma_0$, from
the essential condition of warm inflation $T>H$ which establishes
a lower bound, and the condition of the weak regime i.e.,
$R<1+3gH\dot{\phi}$ which gives an upper bound. Also, we have
observed that the tensor to
 scalar ratio results $r\sim 0$ and the consistency relation $r=r(n_s)$ does not
 add a new constraint on the parameters  during this regime. For
 the case of the strong regime ($1<R+3gH\dot{\phi}$)  we have found
 the constraints on the parameters, only from the condition
 $1<R+3gH\dot{\phi}$ (or strong regime) which gives a lower limit on the parameters.
 In this sense, the essential condition of warm inflation and the
 consistency relation do not give information on the constraints during this regime.

 We mention that the G-warm inflation models are less
restricted than analogous pure warm inflation and that pure
generalized $G$-inflation due to the introduction of a new
parameter; $g_0$ during warm inflation and $\Gamma_0$ in the case
of standard  $G-$inflation. The inclusion of these parameters
gives us a freedom  that allows us to change the pure warm or pure
generalized G inflationary scenarios by simply  modifying the
corresponding values of the parameters. We conclude with some
comments concerning the way to distinguish our model and pure warm
inflation  or pure generalized G-inflation. In our model,
beginning with the background dynamics we noted that from
Eq.(\ref{dphi}) that the velocity of the scalar field (or
equivalently the kinetic term) is much smaller than the standard
warm inflation. Also, in particular for the exponential potential,
we noted that the velocity $\dot{\phi}$ in the case of the pure
generalized G-inflation becomes independent of the potential,
since $\dot{\phi}^2\propto \alpha/g$, in contradistinction to
G-warm and for values of $g\,M_p^3\gg 1$, the velocity is much
smaller than our model. This suggests that the number of e-folds
$N$ at the end inflation during the model of pure generalized
G-inflation is much bigger than our model and than during standard
warm inflation. From the point of view of the perturbative
dynamics we have found that in our model the consistency relation
$r(n_s)\sim 0$, and is similar to the result obtained for the case
of pure warm inflation, see Ref.\cite{Hall:2003zp}. However, one
may expect that in the case of the pure generalized G-inflation
the consistency relation could be  much bigger than in our
model\cite{289}.

\begin{acknowledgments}
This work was supported by  PUENTE Grant DI-PUCV N$_0$
123.748/2017.
\end{acknowledgments}

%\section{appendix}
%\begin{appendix}
%\appendix
%\section{}

%\begin{acknowledgments}
%R.H. was supported by Comisi\'on Nacional de Ciencias y
%Tecnolog\'ia of Chile through FONDECYT Grant N$^{0}$ 1130628 and
%DI-PUCV N$^{0}$ 123.724.
%\end{acknowledgments}

%\\\\\\\\\\\\\\\\\\\\\\\\\\\\\\\\\\\\\\\\\\\\\\\\\\\\\\\\\\\\\\\\\\\\\\\

\end{document}